\documentclass[a4paper,12pt]{article}

\usepackage{amsmath,amsfonts,amssymb,epsfig}

\numberwithin{equation}{section}
\usepackage{cite}

\def\bra{\langle}
\def\ket{\rangle}

\def\tr{\mathrm{tr}}
\def\ap{\alpha^{\prime}}

\def\beq{\begin{equation}}
\def\eeq{\end{equation}}

\def\2b2[#1,#2][#3,#4]{\left( \begin{array}{cc} #1 & #2 \\ #3 & #4 \end{array} \right)}
\def\3b3[#1,#2,#3][#4,#5,#6][#7,#8,#9]{\left( \begin{array}{ccc} #1 & #2 #3 \\ #4 & #5 & #6\\#7&#8&#9\end{array} \right)}

\newcommand{\C}[1]{\mathcal{#1}}

\def\ov{\overline}

\author{K.~Benakli\footnote{kbenakli@lpthe.jussieu.fr}\ \    and M.~D.~Goodsell\footnote{goodsell@lpthe.jussieu.fr}}
\date{}
\title{Dirac Gauginos and Kinetic Mixing}
\begin{document}
\maketitle
\vspace{-1cm}
\begin{center}
\emph{Laboratoire de Physique Th\'eorique et Hautes Energies, \\  CNRS, UPMC Universit\'e Paris 06\\
Boite 126, 4 Place Jussieu, 75252 Paris cedex 05, France}
\end{center}
\abstract{We present formulae for the calculation of Dirac gaugino masses at leading order in the supersymmetry breaking scale using the methods of analytic continuation in superspace and demonstrate a link with kinetic mixing, even for non-abelian gauginos. We illustrate the result through examples in field and string theory. We discuss the possibility that the singlet superfield that gives the $U(1)$ gaugino a Dirac mass may be a modulus, and some consequences of the D-term coupling to the scalar component. We give examples of possible effects in colliders and astroparticle experiments if the modulus scalar constitutes decaying dark matter. }

\section{Introduction}

While non-observation  of light charginos implies  large  gaugino masses,  there are  no indications of their nature. Only Majorana type gauginos are allowed  in the Minimal Supersymmetric Standard Model (MSSM), and as a consequence they have drawn most of the attention in the literature.  However, beyond any theoretical prejudice, both Dirac and Majorana type masses are permitted: they are both soft,  and they both  address the problem of electroweak-Planck scale hierarchy. Therefore,  Dirac gaugino extensions are an alternative and some of their phenomenological   features have been explored (see for example \cite{Fayet:1978qc, Polchinski:1982an, Hall:1990hq, Fox:2002bu, Nelson:2002ca,Antoniadis:2005em,Antoniadis:2006eb,Antoniadis:2006uj,Hsieh:2007wq,Kribs:2007ac,Amigo:2008rc,Marques:2009yu,Blechman:2009if,Benakli:2008pg,Belanger:2009wf,Choi:2008pi,Plehn:2008ae,Choi:2008ub,Kramer:2009kp,Nojiri:2007jm}).

In section 2, we show how to extract Dirac gaugino masses in perturbative models using the methods of analytical continuation in superspace, following the work for the Majorana case in \cite{Giudice:1997ni, ArkaniHamed:1998kj,Hisano:1997ua}. Treating bare couplings  as spurion superfields,  some quantities of phenomenological interest, such as the Majorana gaugino masses, have been shown to be easily extracted from one-loop renormalisation group equations, using regulators that  preserve supersymmetry, threshold matching and analytic continuation. This  a very useful method  as it allows to extract information that would otherwise require  more involved computations of one and two-loop Feynman diagrams. We present its extension for the Dirac gaugino mass computation and provide explicit compact formulae for both  cases of  (potentially but not necessarily $R$-symmetric) $F$-term breaking and $D$-term breaking.

Section 3 points out the peculiarities of $U(1)$  cases. On one hand, for $D$-term breaking, the use of the method of analytical continuation requires the  computation of the coupling of possible abelian kinetic mixing.  On the other hand, the necessary additional adjoint representation (which we denote {\bf  DG-adjoint})  is just a singlet. Such fields are often considered as minimal in extensions for beyond the MSSM,  and they  are expected to be present in many models arising from string or higher dimensional theories. They can appear as light moduli corresponding for instance to some free  deformations of the internal space geometry. We will  point out some new model building opportunities, and  phenomenological implications,  allowed for these singlets (moduli)   by  new interactions, as a consequence of the presence of Dirac gaugino masses.

Section 4 illustrates the explicit use of the method of analytical continuation in order to compute Dirac gaugino masses in an explicit string example.
 We retrieve the result of \cite{Antoniadis:2006eb} in a simpler manner than the calculation  of the full amplitude.

For completeness, section 5  briefly reviews that the appearance of Dirac gaugino masses  can be associated with  gauge symmetry instead of  supersymmetry  breaking. In particular we demonstrate their form for a $U(1)$ breaking  with a St\"uckelberg mechanism.   As conclusions, we summarize our results in section 6.

\section{Dirac Gaugino Masses at Leading Order}
\label{LEADING_SECTION}

In this section we will describe the use of analytic continuation in superspace to calculate Dirac gaugino masses at leading order in the supersymmetry breaking scale. We will show that there is a link with kinetic mixing, which we shall first briefly discuss; for early work on the subject see \cite{Holdom:1985ag,Okun:1982xi,delAguila:1988jz,
Babu:1997st}, for some recent phenomenological applications see \cite{Feldman:2007wj,
Ibarra:2008kn,
ArkaniHamed:2008qn,Morrissey:2009ur},  and for string theory discussion see \cite{Dienes:1996zr,Abel:2003ue,Abel:2006qt,Abel:2008ai,Goodsell:2009xc}.

\subsection{Kinetic Mixing in Supersymmetric Theories}

For holomorphic normalisation of the gauge kinetic terms, the Lagrangian density including a kinetic mixing term is given by
\beq
\mathcal{L} \supset \int d^2 \theta  
\frac{1}{4 (g_h)^2} W^\alpha W_\alpha + \frac{1}{4(g^\prime_h)^2} W^{\prime\,\alpha} W^\prime_\alpha
- \frac{1}{2}\chi_h W^\alpha W^\prime_\alpha ,
\label{KMLag}\eeq
where $W_\alpha, W^\prime_\alpha$ are the field strength superfields for the two U(1) gauge fields and $g_h, g_h^\prime,\chi_h$ are holomorphic quantities that run only at one loop. The existence of states charged under the two  $U(1)$s leads, at one loop, to the mixing term as discussed in  \cite{Holdom:1985ag,Okun:1982xi,delAguila:1988jz,
Babu:1997st}. The holomorphic kinetic mixing $\chi_h$ can be easily calculated. The propagators of the exchanged states involve the mass matrix  elements $\C{M}$ which  can be read from the superpotential $W$, as $\C{M} \equiv W_{ij}$ is the \emph{fermionic} mass matrix (the mass matrix for the bosons is given by $W_{ik} \ov{W}_{kj}$). We can write then 
\beq
\chi_h = -\frac{1}{8\pi^2} \tr \bigg( Q Q' \log \C{M}/\mu\bigg),
\label{ChiH}\eeq
where $Q, Q'$ are the charge operators for the two $U(1)$s. Interestingly for $\mathrm{tr}(Q Q') = 0$ this is independent of $\mu$. It is of course possible to diagonalise these gauge kinetic terms by themselves. This will lead to the appearance of new interaction terms, as explained for example in  \cite{Dienes:1996zr}.
To then find the physical kinetic mixing parameter, that is, the parameter $\chi$ in the canonical  Lagrangian density
\beq
\C{L}_{\mathrm{canonical}} \supset \int d^2 \theta 
\frac{1}{4} W^\alpha W_\alpha + \frac{1}{4} W^{\prime\,\alpha} W^\prime_\alpha
- \frac{1}{2} \chi W^\alpha W^\prime_\alpha,
\eeq
we can apply the techniques of \cite{ArkaniHamed:1997mj}; we rescale the vector superfields from the holomorphic basis (where the gauge kinetic terms are as above) to the canonical basis, $V \rightarrow g V$. In fact, the analysis used there follows through exactly for the kinetic mixing, with the result
\beq
\frac{\chi}{g g'} = \mathrm{Re}(\chi_h) + \frac{1}{8\pi^2} \mathrm{tr}\bigg( Q Q^\prime \log Z \bigg),
\label{new1}\eeq
where $Z$ is the wavefunction renormalisation matrix of chiral superfields, and $g, g'$ are now the physical gauge couplings given by
\beq
g^{-2} = \mathrm{Re} [g_h^{-2}] - \frac{1}{8\pi^2} \mathrm{tr} \bigg( Q^2 \log Z \bigg) 
\label{PhysicalG}\eeq
and similarly for $g'$. In supergravity theories, gauge coupling expressions are augmented by Kaplunovsky-Louis \cite{Kaplunovsky:1994fg,Kaplunovsky:1995jw} terms 
\beq
+\frac{1}{16\pi^2} \mathrm{tr} ( Q^2) \kappa^2 K
\label{KaplunovskyLouis}\eeq
and a similar term for $g'$, where $\kappa^2=8\pi/M_{Pl}^2$ for $M_{Pl}$ the Planck mass, and $K$ the total K\"ahler potential of the theories. It can be shown that the kinetic mixing term above is in turn augmented by 
\beq
-\frac{1}{16\pi^2} \mathrm{\tr} ( Q Q' )\kappa^2 K .
\label{KaplunovskyLouisKM}\eeq

\subsubsection{Magnetic Mixing}

Note in addition that the holomorphic kinetic mixing parameter also contains information about $\theta$-angle mixing. Defining $\mathrm{Im} \left( \frac{1}{g_h^2} \right) = \frac{i \theta}{8\pi^2} $ and similarly for $\theta'$, we have 
\begin{align}
\int d^2 \theta & 
\frac{1}{4 (g_h)^2} W^\alpha W_\alpha+ \frac{1}{4(g^\prime_h)^2} W^{\prime\,\alpha} W^\prime_\alpha
- \frac{1}{2}\chi_h W^\alpha W^\prime_\alpha + c.c. \nonumber \\
\supset&  - \mathrm{Re}(\frac{1}{4 (g_h)^2}) F_{\mu \nu}F^{\mu \nu} - \mathrm{Re}(\frac{1}{4(g^\prime_h)^2})  F_{\mu \nu}^\prime F^{\prime\, \mu \nu} +  \frac{1}{2}\mathrm{Re}(\chi_h) F_{\mu \nu}F^{\prime, \mu \nu} \nonumber \\
& + \frac{\theta}{16\pi^2} F_{\mu \nu}\tilde{F} ^{\mu \nu} + \frac{\theta^\prime}{16\pi^2} F_{\mu \nu}^\prime \tilde{F} ^{\prime\,\mu \nu} - \frac{\theta_M}{8\pi^2} F_{\mu \nu} \tilde{F} ^{\prime\,\mu \nu}
\end{align}
where $\theta_M$ is a ``magnetic-mixing'' angle \cite{Brummer:2009cs,Bruemmer:2009ky}. We then find a ``physical'' magnetic mixing in analogy to the above; the resulting canonical Lagrangian density is 
\begin{align}
\C{L}_{\mathrm{canonical}} \supset \int d^2 \theta& 
\frac{1}{4} g^2 \bigg[  \frac{1}{g_h^2} - \frac{1}{8\pi^2} \mathrm{tr} \bigg( Q^2 \log Z \bigg) \bigg] W^\alpha W_\alpha \nonumber \\
&+ \frac{1}{4} (g')^2 \bigg[ \frac{1}{(g_h^\prime)^2} - \frac{1}{8\pi^2} \mathrm{tr} \bigg( (Q')^2 \log Z \bigg) \bigg] W^{\prime\,\alpha} W^\prime_\alpha \nonumber \\
&- \frac{1}{2} g g' \bigg[ \chi_h + \frac{1}{8\pi^2} \mathrm{tr} \bigg( Q Q' \log Z \bigg) \bigg] W^\alpha W^\prime_\alpha \nonumber \\
+c.c.&
\label{RescaleTheta}\end{align}
where the physical couplings $g, g'$ are given above in (\ref{PhysicalG}); (\ref{RescaleTheta}) then has straightforward modifications in the case of supergravity of adding the corresponding terms from (\ref{KaplunovskyLouis}), (\ref{KaplunovskyLouisKM}) to the square brackets. 
Given that $Z$ is a real function (and, for supergravity, so is $K$), we find 
\begin{align}
\C{L}_{\mathrm{canonical}} \supset  + \frac{g^2 \theta}{16\pi^2} F_{\mu \nu}\tilde{F} ^{\mu \nu} + \frac{(g')^2 \theta^\prime}{16\pi^2} F_{\mu \nu}^\prime \tilde{F} ^{\prime\,\mu \nu} - \frac{g g' \theta_M}{8\pi^2} F_{\mu \nu} \tilde{F} ^{\prime\,\mu \nu}.
\end{align}
Since $\chi_h$ runs only at one loop, the same is true of $\theta, \theta', \theta_M$; in fact the latter is only generated at one loop and is given by
\beq
\theta_M = - \mathrm{Im} \bigg[ \tr \bigg( Q Q' \log \C{M}/\mu\bigg)\bigg].
\eeq

\subsection{F-Terms}

There are two  sources of Dirac gaugino masses at leading order in the supersymmetry breaking parameter:  $F$-terms or $D$-terms. The first applies only to $U(1)$ gauginos; if the visible gaugino $W_\alpha = \lambda_\alpha + ...$ kinetically mixes with a hidden gaugino $W^\prime_\alpha=\lambda_\alpha^\prime + ...$ with coupling
\beq
\C{L} \supset -\frac{1}{2} \int d^2 \theta \ W^\alpha W^\prime_\alpha \chi(S) 
\eeq
where $\chi(S)$ is a holomorphic function of chiral superfields, including a (pseudo)modulus $S$. If $S$ develops an $F$-term, this leads to a Dirac gaugino mass term
\beq
\C{L} \supset -\frac{1}{2} (\partial_S \chi(S)) F_S \lambda^\alpha \lambda^{\prime}_\alpha .
\label{FTermDiracMass}\eeq

\subsection{D-Terms}
\label{DTerms}

For $SU(N)$ gauginos, the only gaugino mass term that can be generated at leading order in the supersymmetry breaking parameter is via $D$-terms; $D$-terms may of course also induce Dirac masses for $U(1)$ gauginos. They occur via the operator
\beq
\int d^2 \theta \frac{\kappa}{\Lambda}\tr(W^{\prime \alpha} W_{\alpha} \mathcal{X}),
\label{masterop}
\eeq
where $\C{X} = X + \sqrt{2}\theta \chi + (\theta \theta) F_X + ...$ is a chiral field in the adjoint representation of the same gauge group as $W_\alpha$. Here, we have parametrised the $D$-term  using a spurion $W^\prime_\alpha$, with the following two options:
\begin{enumerate}
\item $\bra W^\prime_\alpha \ket = \theta_\alpha D'$ is from a $U(1)$ gauge field (hidden \emph{or hypercharge})
\item $W^\prime_\alpha =  W_\alpha$, and  $\bra W_\alpha \ket = \theta_\alpha T^A D^{A}$ is a vev for the $SU(N)$ field 
\end{enumerate}
We shall consider both possibilities below.

\subsubsection{Using $U(1)$ D-term}

Consider the first option $\bra W^\prime_\alpha \ket = \theta_\alpha D'$. The expansion in components of (\ref{masterop})
\begin{align}
\frac{\kappa}{\Lambda}\int d^2 \theta W^{\prime \alpha} W_{\alpha} X \supset&\frac{\kappa}{\Lambda}\int d^2 \theta \lambda^{\prime \alpha} \frac{1}{2} i (\sigma^\mu \ov{\sigma}^\nu)_\alpha^\beta F_{\mu \nu} \theta_\beta \sqrt{2} \theta \psi + \theta^\alpha D^\prime \lambda_\alpha \sqrt{2} (\theta \psi) \nonumber \\
& - \frac{1}{4} \epsilon^{\gamma \alpha}(\sigma^\mu \ov{\sigma}^\nu)_\alpha^\beta F_{\mu \nu}^\prime \theta_\beta (\sigma^\rho \ov{\sigma}^\kappa)_\gamma^\delta F_{\rho \kappa} \theta_\delta \\
=&-\frac{\kappa}{\Lambda}\frac{1}{\sqrt{2}} \lambda^{\prime \alpha} \frac{1}{2} i (\sigma^\mu \ov{\sigma}^\nu)_\alpha^\beta F_{\mu \nu} \psi_\beta -\frac{\kappa}{\Lambda} \frac{1}{\sqrt{2}} D^\prime (\lambda \psi) - \frac{\kappa}{\Lambda}\frac{1}{2} F^{\prime \mu \nu} F_{\mu \nu} X  \nonumber
\end{align} 
shows that supersymmetry relates the Dirac gaugino mass and the kinetic mixing; a vev for $X$ generates a kinetic mixing operator. Indeed by expanding (\ref{KMLag}) we see that we generate the operator (\ref{masterop}) by
\beq
\frac{\kappa}{\Lambda} = - \frac{1}{2}\partial_X (\chi(X))\bigg|_{X=\bra X \ket},
\label{new2}
\eeq
where we differentiate with respect to the bosonic component and set it equal to its vacuum expectation value. Then we can derive the value of the corresponding Dirac gaugino mass as:
\beq
m_D =  - \frac{1}{2} \frac{D^\prime}{\sqrt{2}} \partial_X \chi(X)|_{X=\bra X \ket} = - \frac{1}{2} \frac{g g'}{8\pi^2}\frac{D^\prime}{\sqrt{2}} \partial_X \tr \bigg(Q Q' \log \C{M}(X)/\mu \bigg)\bigg|_{X=\bra X \ket}.
\eeq
For a non-abelian group we compute the mass for the $U(1)$ generators in the Cartan subalgebra and deduce from gauge invariance the induced mass. We can write $W_{\alpha} = W^I_\alpha T^I$, and the kinetic mixing term is
\beq
-\frac{1}{2}\int d^2 \theta W^{\prime\alpha} W^I_{\alpha} \chi(X^I)
\eeq
where we should find
\begin{align}
\chi(X^I) = \frac{g g'}{8\pi^2} \tr \bigg( Q' R(T^I) \log \C{M}(X^I)/\mu\bigg) 
\end{align}
where $R(T^I)$ denotes the representation of $T^I$ (i.e. fundamental or antifundamental); this appears as the ``charge'' of the new $U(1)$, which couples via $\tr (\Phi^\dagger 2gV^I R(T^I) \Phi)$, c.f. $\Phi^\dagger 2 g Q V \Phi$ for a $U(1)$. This then yields an operator
\beq
- \frac{1}{2} \int d^2\theta W^{\prime\alpha} W^I_{\alpha} X^I \partial_{X^I} \chi (X^I) = - \frac{1}{2} \int d^2 \theta\  2 W^{\prime\alpha} \tr (W_{\alpha} X) \partial_{X^I} \chi(X^I)
\eeq
where we normalise the generators to $\mathrm{tr}(T_I T_J) = \frac{\delta^{IJ}}{2}$, and thus gaugino masses of
\beq
m_D = - \frac{1}{2} \frac{D^\prime}{\sqrt{2}} \frac{g g'}{8\pi^2} \partial_{X^I} \tr \bigg( Q' R(T^I) \log \C{M}(X^I)/\mu\bigg)\bigg|_{X^I=0} .
\eeq
Note here that the vacuum expectation value of $X^I$ is zero since we are assuming that the gauge group is unbroken. We should take $T^I$ to lie in the Cartan subalgebra of the group, and thus for SU(N) a convenient representation is $\pm\frac{1}{2} \mathrm{diag}(1,-1,0,0,0,...,0)$ where the upper (lower) sign is for the (anti)fundamental representation.

\subsubsection{Using $SU(N)$ D-term}

Considering the possibility that $\bra W_\alpha \ket = \theta_\alpha T^A D^{ A}$, we find Dirac gaugino masses via 
\begin{align}
\int d^2 \theta \frac{\kappa}{\Lambda} \tr(W^{\alpha} W_{\alpha} X) &\supset -\frac{1}{\sqrt{2}} \frac{\kappa D^{ A}}{\Lambda}\tr(\{T^A, T^B\} T^C) (\lambda^B \chi^C) \nonumber \\
&\supset -\frac{1}{\sqrt{2}} \frac{\kappa D^{ A}}{\Lambda} \C{A}^{ABC} (\lambda^B \chi^C)
\end{align}
where $\C{A}^{ABC}$ is the anomaly coefficient; this is zero for $SU(2)$, thus we cannot generate a mass for it this way at this order, but for $SU(3)$ we have $\C{A}^{ABC} = \frac{1}{2} d^{ABC}$.
Using the same technique as for the case of $U(1)$ D-term, we can determine the coefficient to be 
\beq
\frac{\kappa }{\Lambda} = - \frac{1}{2} \frac{g g'}{8\pi^2} \partial_{X^I}\tr \bigg(R(T^I T^I) \log \C{M}/\mu\bigg)\bigg|_{X^I=0} .
\eeq

\section{Visible moduli and  phenomenological aspects}

This section contains a few comments on supersymmetric extensions of the Standard Model where  Dirac gaugino masses are relevant and could  play an important role.  We consider then that the model is an extension of the Minimal Supersymmetric Standard Model by adjoint representations, a singlet for $U(1)$, a triplet for $SU(2)$ and an octet for $SU(3)$. The possible presence of other  extra states (singlets, extra Higgs doublets, ...)  is irrelevant  for the present discussion, and the minimal content with the corresponding Lagrangian has been described in detail in \cite{Belanger:2009wf}. Obviously, we look for situation where Dirac masses should not represent  a negligible perturbation around possible Majorana ones. The relative sizes depend on the details of the supersymmetry breaking and mediation sectors. We can consider three possible outputs of the supersymmetry breaking  where one needs to care about the Dirac contributions to gaugino masses. We consider, for simplicity,  a single gaugino:

\begin{itemize}

\item The first and most important case is when the Dirac masses are much bigger than the Majorana ones. Majorana masses can be vanishing (or  negligibly small)  if for instance the supersymmetry breaking preserves $R$-symmetry (or breaks it very weakly).

\item The second case consists in one large and one vanishing Majorana masses. For instance, if the extra adjoint state,  denoted DG-adjoint, has a Majorana mass $M$ while the the gaugino one vanishes, the resulting lightest Majorana mass  after diagonalisation is of order $m_D^2/M \ll m_D \ll M$ leading to a hierarchically lighter gaugino through a see-saw mechanism. 

\item The third case is when the two Majorana masses are degenerate with a common value $M$. After diagonalisation, the Dirac mass induces a splitting of order $2 m_D$ between the two states which can be relevant. For example, if $m_D$ if small there could be important coannihilations between the two states in early universe, or they might lead to interesting signatures at colliders.

\end{itemize}

Of course, one needs to extend the supersymmetric Standard Model field content by the DG-adjoints representations that couple to the gauginos. However, this is a very minimal extension for the the case of $U(1)$  as the necessary adjoint is just a singlet superfield $\mathbb{S}=S + \sqrt{2} \theta \chi_S + \cdots$, where $S= \frac{1}{\sqrt{2}}(S_R+i S_I)$ is a complex scalar field\footnote{The singlet-gaugino coupling could be forbidden by some discrete symmetries, as for the singlet of the NMSSM, for example.}. Such  singlets are introduced in many extensions of the Standard Model. In particular, we would like to discuss here the possibility that this can be identified with a generic modulus field.

The Dirac gaugino mass term appears in the action as
\begin{eqnarray}
 \int d^4x d^2\theta    \sqrt{2} \textbf{m}^\alpha_{D} \mathbf{W}_{\alpha} \mathbb{S}  
\label{Newdiracgauge}
\end{eqnarray}
where  $\mathbf{W}_{j \alpha} $ are the corresponding field strength superfields associated to $U(1)_Y$, $ SU(2)$ and $ SU(3)$  for $j=1, 2, 3$ respectively. We have introduced  a spurion superfield:
\begin{eqnarray}
\textbf{m}_{\alpha D} = \theta_\alpha m_{D}.
\end{eqnarray}
The above spurion superfield can be written as  $\textbf{m}^\alpha_{D} = -\frac{1}{4} \bar {D}\bar{ D}D_\alpha \mathbf{X} $, where $\mathbf{X} = \mathbf{V'}/ \Lambda$ is a vector superfield with a suppression scale $\Lambda$, typically of order of the Planck scale $M_{Pl}$ if $\mathbb{S}$ is a modulus\footnote{ Although we consider in our examples  a suppression by $M_{Pl}$, one could consider moduli parametrising flat directions with coupling to matter suppressed by smaller scales as the string of GUT scale, due to presence of an internal space compactification volume factor for example.}. This leads  to the operator 
\begin{eqnarray}
 \int d^4x d^2\theta    \frac{1}{4\Lambda} \mathbf{W'}^{\alpha} \mathbf{W}_{\alpha} \mathbb{S}  &=& -\int d^4x  \frac{<D'>}{\Lambda}\lambda^{\alpha} \chi_{S\alpha} + ... \nonumber \\ &=& -\int d^4x m_D \lambda^{\alpha} \chi_{S\alpha} + ...
\label{diracgaugeI}
\end{eqnarray}

On the other hand, a non-vanishing  $F$-term origin appears at second order in $F$, and is identified by writing $\mathbf{X} = 2 \mathbf{\Sigma^\dagger\Sigma}/ \Lambda^3$  with $\mathbf{\Sigma} = \theta\theta F$, giving 
\begin{eqnarray}
 -\int d^4x d^2\theta    \frac{1}{4\Lambda^3} \mathbb{S} \mathbf{W}^{\alpha}  \bar {D}\bar{ D}D_\alpha (\mathbf{\Sigma^\dagger\Sigma})   &=& -\int d^4x  \frac{|F^2|}{\Lambda^3}\lambda^{\alpha} \chi_{S\alpha} + ... \nonumber \\ &=& -\int d^4x m_D \lambda^{\alpha} \chi_{S\alpha} + ...
\label{diracgaugeII}
\end{eqnarray}

In order to identify the suppression scale $\Lambda$, let us calculate this amplitude for the case that the spurion $\Sigma$ differs from the modulus $\mathbb{S}$, and couples to messenger fields via a superpotential interaction of
\beq
W \supset \sum_{i,j}  \frac{1}{2} W_{ij} (S) \phi_i \phi_j +  \frac{1}{2} W_{\Sigma ij}(S) \Sigma \phi_i \phi_j.
\eeq
We need only compute a three-point interaction involving the auxiliary fields $F_\Sigma, F_{\Sigma}^\dagger$ and either the auxiliary $D$ or a gauge boson. Since we compute these in the supersymmetric limit (we are only interested in the leading order SUSY breaking term) we can work in a diagonal mass basis (i.e. $W_{ij} (S)W_{jk}^{\star}(\ov{S}) = m_i^2 \delta_{ik}$) such that the gauge current is also diagonal; $\mathcal{L} \supset g D \sum_i q_i \phi_i^\dagger \phi_i$, where $q_i$ is the charge under the $U(1)$. Then the result is
\beq
\frac{1}{\Lambda^3} = -\partial_S \bigg[\frac{1}{16\pi^2} \sum_{i,j}  q_j f_{ij} W_{\Sigma i j } ( W_{\Sigma i j })^\dagger \bigg]
\eeq
where 
\beq
f_{ij} \equiv \left\{ \begin{array}{cc} \frac{m_j^2 - m_i^2 + m_i^2 \log m_i^2/m_j^2}{(m_j^2 - m_i^2)^2} & m_i^2 \ne m_j^2 \\\frac{1}{2m_i^2} & m_i^2 = m_j^2 \end{array} \right.
\eeq
For identical masses $m_i^2 = M^2$ the result then simplifies to 
\beq
\frac{1}{\Lambda^3} = -\frac{1}{16\pi^2} \partial_S \bigg[ \frac{1}{2 M^2} \sum_{i,j} q_j W_{\Sigma i j } ( W_{\Sigma i j })^\dagger \bigg].
\eeq

In order to be more explicit, as a toy example, let us consider a simple model with four messenger fields $\phi_i, i=1..4$ having charges $(-1)^i$  under the $U(1)$, a modulus $S$ and superpotential
\beq
W = M \bigg[ e^{-\frac{a \mathbb{S}}{M_{Pl}}} \phi_1 \phi_4 + e^{\frac{a \mathbb{S}}{M_{Pl}}}\phi_3 \phi_2 \bigg] + \Sigma \phi_1 \phi_2.
\label{PositiveModel}\eeq
We then find that to order $|F|^2$
\begin{align}
|m_D |=& \frac{1}{16\pi^2} \frac{a|F|^2}{M^2 M_{Pl}} \frac{e^{-\frac{a (S+\ov{S})}{M_{Pl}}}}{(1-e^{-2\frac{a (S+\ov{S})}{M_{Pl}}})^3} \nonumber \\
&\times \bigg[-4 +  4e^{-4\frac{a (S+\ov{S})}{M_{Pl}}} + 2\frac{a (S+\ov{S})}{M_{Pl}}(1+3e^{-2\frac{a (S+\ov{S})}{M_{Pl}}})^2 \bigg].
\end{align}
Note that the above can be replaced by a renormalisable model by taking the linear approximation of the exponentials and defining $y \equiv M/M_{Pl}$, giving $|m_D| = \frac{1}{16\pi^2} \frac{a y|F|^2}{M^3} \frac{1}{3}$. This model also generates positive mass-squareds for the scalars \cite{Benakli:2008pg,Amigo:2008rc,Blechman:2009if}, and we thus expect this to be true for the generalisation (\ref{PositiveModel}); the moduli are therefore stabilised at a small value.

Denoting the observable supersymmetry breaking scale as $M_{soft}$, to be of order of the TeV, and taking $m_D \sim M_{soft}$  constraints  the supersymmetry breaking vevs as  
\begin{eqnarray}
<D'> \sim M_{soft}  \Lambda \qquad {\rm or} \qquad F \sim \Lambda^{3/4} M_{soft}^{1/4}
\label{scalesI}
\end{eqnarray}
which become for the modulus example cases:
\begin{eqnarray}
<D'> \sim M_{soft}  M_{Pl} \qquad {\rm or} \qquad F \sim M^{1/2} M_{soft}^{1/4} M_{Pl}^{1/4}
\label{scalesII}
\end{eqnarray}
with $M$ the mass of messengers mediating the supersymmetry breaking.

One important implication of the Dirac gaugino masses is that its soft nature requires a new interaction between the singlet scalar and the original (before including Dirac gaugino mass) D-term $D_1^{(0)}$ of $U(1)$. The scalar potential contains terms such as:
\begin{eqnarray}
V  &\supset &  2 m^2_{1D}  S^2_R -     2 m_{1D}  S_R    D_1^{(0)}  +  \frac{1}{2}   D_1^{(0)2} \nonumber \\
D_1^{(0)} &=&  - g \sum_{j} Y_j \varphi_j^* \varphi_j 
\label{newinter}
\end{eqnarray}
where the sum in the second equation holds on all charged scalar fields, $Y_j$ being their charge and $g$ the $U(1)$ coupling constant. This shows that the real part of the singlet $S$ has acquired new interactions of dimension one. In particular, we would like to consider the case of ${S}$ being a modulus field  and discuss some implications of the interactions presented. In the supersymmetric limit, a generic modulus $S$ couples to matter only gravitationally, thus with non-renormalisable operators suppressed by $M_{Pl}$, as assumed in the $D$ and $F$-term supersymmetry breaking examples above. The new feature is that now the supersymmetry breaking has ``automatically'' generated an interaction that is no longer suppressed by $M_{Pl}$, but is dimensionful and will lead to suppression by the ratio $m_D/M_S$ where  $M_S$ is the (real part) modulus  mass. The  ratio $m_D/M_S$  does not need to be small. We would like to illustrate the possible use in model building of (\ref{newinter}) with phenomenological implications, without going into explicit details which are beyond the scope of this work. For this purpose, we will consider two different cases.

\subsection*{Bino - modulino mixing}
First, let us identify in (\ref{Newdiracgauge}) $\mathbf{W}_{\alpha}$ with the standard hypercharge $U(1)_Y$. The corresponding  gaugino, the bino, has a soft supersymmetry breaking Dirac mass through mixing with a modulino field. This is a particular case of situation discussed for example in \cite{Belanger:2009wf}, and all the analysis there applies here. 

There are two implications stressed in \cite{Belanger:2009wf}:
\begin{itemize}
\item The singlet $S_R$ mixes with the neutral Higgs. It modifies the production and decays of the light and/or heavy Higgs and can be be produced associated with them.

\item  The interaction  (\ref{newinter}) implies that $S_R$ can be produced at colliders and will decay to sparticles or standard model states depending on the region kinematically allowed by the spectrum.  For example, consider that $S_R$ mass is lighter than the sfermion ones. While the MSSM Higgs has couplings to matter states that are given by Yukawa couplings, the singlet field $S_R$ couples directly only to scalars (squarks, sleptons and Higgses). These couplings are  all proportional to the gauge couplings. As we consider the case where $S_R$ is lighter than the squarks and sleptons, these cannot be the final states, but should appear as virtual intermediary states. Taking into account the available phase space and the strength of sfermion couplings to final states, the   gluons represent, obviously, the most important final states channel. It can be emitted  either by Higgs bosons, squarks or sleptons and will   decay by  producing vector bosons and pairs of leptons or quarks. The main decay channel  is into two gluons through triangle diagrams involving intermediate squarks with masses $M_{\tilde q_i}$. The decay width behaves as
\begin{eqnarray}
\Gamma \sim \alpha \alpha_s^2 \frac{M_D^2}{M_S} f(M_S,M_{\tilde q_i}) \sim \alpha \alpha_s^2 \frac{M_D^2 M_S^3}{M^4_{\tilde q}}
\label{lifetime0}
\end{eqnarray} 
where $\alpha_s$ is the strong coupling and  $ M^4_{\tilde q}$ is defined as an average squark mass splitting. For all masses around the electroweak scale, the lifetime is small and implies a decay inside the collider. It would be interesting to see if it possible to isolate the corresponding signals from the Standard Model background.

\end{itemize}
 
\subsection*{Hidden $U(1)$ - modulino mixing}

Let us consider instead a modulus $S_R$ that couples only to give a Dirac mass to a hidden $U(1)_H$ gaugino.  We suppose also that there are some pairs of heavy states $\Phi_i$ with supersymmetric masses of order $M_{\Phi}$ which carry charges under both  $U(1)_H$ and the Standard Model gauge group. The  breaking of supersymmetry, in addition to the generation of the Dirac gaugino mass $m_D$ and the modulus mass $M_S$, is assumed to induce a mass splitting $\Delta M^2_{\Phi}=M^2_{\Phi,+}-M^2_{\Phi,-}$ between opposite charged states. We suppose $\sqrt{|\Delta M^2_{\Phi}|} < M_{\Phi}$.

The interaction terms in (\ref{newinter}) allow for instance the decay of $S_R$ into two photons with a life time of order:
\begin{eqnarray}
\tau \sim \frac{2304 \pi^2}{\alpha^2_{em} \alpha_H} \frac{M_{\Phi}^8}{m_D^2 M^3_S |\Delta M^2_{\Phi}|^2}
\label{lifetime}
\end{eqnarray}
where $\alpha_{em}$ and $\alpha_{H}$ are the electromagnetic and  $U(1)_H$ coupling constants, respectively.

A possible application of this is  a scenario trying to identify the modulus $S_R$ as a dark matter candidate.  While usually moduli  interactions are too weak - with annihilation cross section of order $1/ M_{pl}^2$ - to allow such a possibility, the supersymmetry breaking  terms in (\ref{newinter}) lead to extremely enhanced annihilation cross sections roughly of order $\frac{1}{M_s^2}(\frac{m_D}{M_S})^n$ where the power $n$ depends on the precise interaction of the scalars in the $D$-terms of $U(1)_H$. Depending on the thermal history of the hidden sector, it is then possible to engineer a model with $S$ as dark matter component. For $m_D$, $M_S$ and 
$\sqrt{|\Delta M^2_{\Phi}|}$ of order TeV, with $\alpha_{H}\sim 10^{-2}$, a choice of  $ M_{\Phi}$ in the range $10^4-10^5$ GeV leads to a life time of order $10^{26}s$, thus with decay products observable by present time experiments such as PAMELA and FERMI (see for example \cite{Chen:2008yi,Nardi:2008ix,Pospelov:2008rn,Arvanitaki:2008hq}). The states $\Phi_i$ induce also a $U(1)$ kinetic mixing of order $10^{-6}$. Note that the value of the mass range $ M_{\Phi}$ can be lowered to TeV scale by choosing the hidden sector to be extremely weakly coupled, $\alpha_{H}\sim 10^{-10}$.

\section{Type II String Theory Application}

Above we established the link between kinetic mixing and Dirac gaugino masses. In this section we shall apply this to examine when we may obtain Dirac gaugino masses at leading order from type II string theory; kinetic mixing has been discussed in string theory in \cite{Dienes:1996zr,Abel:2003ue,Abel:2006qt,Abel:2008ai} and its holomorphic nature in particular in \cite{Goodsell:2009xc}.

\subsection{F-Terms}

A very simple model for generating a Dirac gaugino mass through F-terms consists of two messenger fields $\phi_1,\phi_2$ of charges $(\pm 1, \mp 1)$ under the visible and a hidden $U(1)$ with a mass term
\beq
W \supset  M f(e^{-\frac{\mathbb{S}}{M_{Pl}}}) \phi_1 \phi_2,
\eeq
where $\mathbb{S}$ denotes some modulus. 
We then find kinetic mixing via (\ref{ChiH}) of
\beq
\chi_h = \frac{1}{4\pi^2} \log \frac{M f(e^{-\frac{\mathbb{S}}{M_{Pl}}})}{\mu} + \frac{i}{8\pi},
\eeq
and therefore Dirac gaugino masses via (\ref{FTermDiracMass}) of
\beq
m_D = -\frac{1}{2} \frac{1}{4\pi^2} \frac{F_S}{M_{Pl}}e^{-\frac{\mathbb{S}}{M_{Pl}}} \frac{f'}{f}\bigg|_{\mathbb{S} = \bra S \ket}. 
\eeq
It would be interesting to identify the modulus (or moduli) $\mathbb{S}$ in explicit string models. However, through analysis of the dimensional reduction of string effective actions we can determine which moduli can contribute and at what order (perturbative or non-perturbative) in perturbation theory.

As discussed in \cite{Goodsell:2009xc,Akerblom:2007uc} in type IIA string theory the holomorphic kinetic mixing is given by
\beq
\chi_{h,\,IIA} = \chi_{IIA}^{\mathrm{P}} (e^{-T_\alpha},y) + \chi_{IIA}^{\mathrm{NP}} (e^{-z^K},e^{-T_\alpha},y)
\eeq
where $y$ are open string moduli, $T_\alpha$ are K\"ahler moduli and $z^K$ are complex structure moduli; $\mathrm{P,NP}$ denote perturbative (one loop only) and non-perturbative contributions respectively. In type IIB, the expression is
\beq
\chi_{h,\,IIB} = \chi_{IIB}^{\mathrm{P}} (z^K,y) + \chi_{IIB}^{\mathrm{NP}} (z^K,e^{-T_\alpha},y).
\eeq
The exponential dependence upon certain moduli above is due to Peccei-Quinn shift symmetries. This dependence is the same as found in the superpotential; interestingly, one method that non-perturbative corrections in the kinetic mixing may be generated is through non-perturbative contributions to the superpotential, as can be seen in our simple example above if we take $f (e^{-\frac{\mathbb{S}}{M_{Pl}}})$ to be of non-perturbative origin.  

From the above we conclude that if we give $F$-terms to only the complex structure moduli in type IIA string theory, or to the  K\"ahler moduli in type IIB,  Dirac gaugino masses at leading order through $F$-terms and gravity mediation can only be generated via non-perturbative effects.

\subsection{D-term Example}

We would now like to illustrate the application of the formulae of subsection (\ref{DTerms}) to the simple explicit string model treated in \cite{Antoniadis:2006eb}. 
Consider an $N=2$ supersymmetric model of intersecting $D6$ branes in type IIA string theory on a six-torus $\mathbb{T}^6 = \mathbb{T}^2 \times \mathbb{T}^2 \times \mathbb{T}^2$. To be fully consistent this should be an orientifold, to cancel $RR$ charges, with an additional orbifold to break the supersymmetry of the bulk to $N=1$ from $N=4$, but for our purposes we can consider just the local structure of two stacks  $a, b$ of $N_a, N_b$  $D6$ branes   parallel but separated by a distance $l$ in the first torus, and intersecting at angles $\alpha_2, \alpha_3$ in the second and third respectively.  
For the supersymmetric state $\alpha_2 + \alpha_3 = 0, \alpha_2 \ne 0$; we can break supersymmetry by deforming this to $\alpha_2 + \alpha_3 = \epsilon$. The masses for the lightest scalars are given by $\ap m^2 = \epsilon + l^2/4\pi^2\ap$, corresponding to a Fayet-Iliopoulos term for the $U(1)$ on the brane that is deformed from its supersymmetric cycle of $\xi = \epsilon/\ap$, which becomes a $D$-term due to the supersymmetric contribution to the mass. The deformation can be effected by deforming the complex structure moduli of the tori (the FI term is implicitly a function of these). Alternatively, we can give an F-term to the K\"ahler modulus of torus one; the K\"ahler modulus $T = T_1 + i T_2$ has vev $\bra T \ket = i \bra T_2 \ket = R_1 R_2 \sin \alpha$ ($R_1, R_2$ are the radii and $\alpha$ is the angle between the radial cycles).

The kinetic mixing between the $U(1)$s supported on branes $a$ and $b$ can be explicitly computed by using (\ref{ChiH}) and plugging in the string states mass formulae. It is given by \cite{Berg:2004ek,Abel:2006yk,Abel:2008ai}
\beq
\chi_{ab} = \frac{1}{4\pi^2} I_{a,b}^{2} I_{a,b}^{3} \bigg[\log \left| \frac{\theta_1 (\frac{i l L}{4\pi^2 \ap}, 
\frac{i T_2}{\ap})}{\eta(\frac{iT_2}{\ap})} \right|^2 - \frac{l^2}{8\pi^3 \ap} 
\frac{L^2}{T_2}\bigg]\, ,
\eeq
where $I_{a,b}^{2,3}$ are the number of intersections between the branes in tori two and three, and $L$ is the length of the branes in that torus. For a rectangular torus one of common radii $R_1 = R_2 = R$, we can write $T_2 = R^2, L= 2\pi R$.

We can use the above expression to illustrate the field theory derivations in section \ref{LEADING_SECTION}. This can be either by giving a Dirac gaugino mass mixing two $U(1)$s supported one on each D-brane, or by mixing with the adjoint fields; each D-brane actually has three adjoints $\Phi^i_{a,b}$ where $i$ denotes an excitation in the $i^{\mathrm{th}}$ torus, but $\Phi^1_{a,b}$ are the states that can mix with the gaugino, since prior to supersymmetry breaking they preserve the same $N=2$ supersymmetry as both branes (the branes individually preserve $N=4$ supersymmetry, but mutually only preserve $N=2$ so the chiral states are only of the lesser amount). 

Firstly, for $N_a = N_b = 1$ (so two $U(1)$ gauge groups) and taking brane $b$ to be the brane carrying a $D$-term by being tilted (so that it is the gauginos of brane $a$ that obtain a mass, with the adjoint $\Phi^1_a$ becoming $X$ in our earlier formulae) the distance $l$ is proportional to the difference in vevs of $\bra \Phi^1_a - \Phi^1_b \ket = \bra \Phi^1_a \ket$. The exact equivalence is determined by examining the DBI action:
\begin{align}
S_a &= 2\pi (4\pi^2\ap)^{-7/2} g_s^{-1} \int_a d^7 x \sqrt{\det g + 2\pi \ap F} \nonumber \\
&=\mathrm{const} + \int d^4 x \frac{1}{4g_{YM}^2} \tr F_{\mu \nu} F^{\mu \nu} + ...
\end{align}
and identifying $\mathrm{Im}(\Phi^1_{a})$ with $A^1_a$, the component in the first torus of the gauge field. This gives a kinetic term for $\Phi^1_{a}$ of
\beq
S_{\Phi_a^1} = \int d^4 x \frac{1}{g_{YM}^2}  D_{\mu} (\Phi^1_{a})^{\dagger} D^{\mu} \Phi^1_{a}.
\eeq
(or $\frac{2}{g_{YM}^2} \tr D_{\mu} \Phi^1_{a} D^{\mu} \Phi^1_{a}$ for $SU(N)$ adjoints).
Now the mass splitting for states due to a separation of $l$ is given by
\beq
m^2 = \frac{l^2}{4\pi^2 (\ap)^2} = \bra (\Phi^1_a - \Phi^1_b)^2 \ket
\eeq
and thus $l = 2\pi \ap \bra \Phi^1_a - \Phi^1_b \ket$. We therefore find 
\beq
\frac{\kappa}{\Lambda} = - \frac{1}{2} \partial_X \chi = - \frac{1}{2} 2\pi \ap \partial_l \chi = - \frac{1}{2} \frac{1}{4\pi^2} \frac{L}{2\pi} I_{a,b}^{2} I_{a,b}^{3} \bigg[ i\frac{\theta_1^\prime (\frac{i l L}{4\pi^2 \ap}, 
\frac{i T_2}{\ap})}{\theta_1 (\frac{i l L}{4\pi^2 \ap}, \frac{i T_2}{\ap})} - \frac{l L}{2\pi T_2} \bigg].
\eeq
The singular behaviour of the above function as $l \rightarrow 0$ is
\beq
\frac{\kappa}{\Lambda} \rightarrow - \frac{1}{2} \frac{1}{4\pi^2} I_{a,b}^{2} I_{a,b}^{3} \frac{2\pi \ap}{l} + ... = - \frac{1}{2} \frac{1}{4\pi^2} I_{a,b}^{2} I_{a,b}^{3} \frac{1}{X}
\eeq
and is a result of having integrated out all modes of mass greater than $l/2\pi \ap$; recall that $\Phi^1_a \equiv X$. In fact this can also be understood from a field theory perspective: suppose that the mixing is mediated by  $I_{a,b}^{2} I_{a,b}^{3}$ pairs of messenger fields $Q, \tilde{Q}$ with superpotential $X Q \tilde{Q}$. Then we find
\beq
\frac{\kappa}{\Lambda} = - \frac{1}{2} \partial_X \chi = - \frac{1}{2} \frac{2}{16\pi^2} I_{a,b}^{2} I_{a,b}^{3}\partial_X \log X^2 = - \frac{1}{2} \frac{1}{4\pi^2} I_{a,b}^{2} I_{a,b}^{3} \frac{1}{X},
\eeq
in agreement with the above. The messengers $Q, \tilde{Q}$ are to be identified with excitations of strings that are stretching between the two stacks of branes.

Similarly for a stack of $N_a$ branes (keeping $N_b=1$), the relevant generator to break to $U(1)\times SU(N-2)$ corresponds to displacing one brane by $dX/4\pi\ap$ and one by $-dX/4\pi\ap$ (in this case, prior to gaining a mass via the St\"uckelberg mechanism, we actually have $U(1) \times U(1)\times U(N-2)$ and our choice will correspond to taking one linear combination of the two $U(1)$ factors). Thus we have 
\beq
\frac{\kappa}{\Lambda} = -\frac{1}{2} \frac{1}{4\pi^2} \frac{L}{2\pi} I_{a,b}^{2} I_{a,b}^{3} \bigg[ i\frac{\theta_1^\prime (\frac{i l L}{4\pi^2 \ap}, 
\frac{i T_2}{\ap})}{\theta_1 (\frac{i l L}{4\pi^2 \ap}, \frac{i T_2}{\ap})} - \frac{l L}{2\pi T_2} \bigg] \rightarrow -\frac{1}{2} \frac{1}{4\pi^2} I_{a,b}^{2} I_{a,b}^{3} \frac{1}{X}
\eeq
This reproduces the result of \cite{Antoniadis:2006eb}  which was obtained through the more tedious direct calculation  of  amplitudes with insertions of the necessary  vertex operators.

\section{Supersymmetric Dirac Gaugino Masses}

While we have discussed Dirac gaugino masses induced by supersymmetry breaking,  we would like to review for the simple example of $U(1)$s that such masses can  be supersymmetric, coming with gauge symmetry breaking, for instance through a St\"uckelberg mechanism (see \cite{Kors:2004ri}).

It is possible for $U(1)$ gauginos to acquire supersymmetric Dirac masses via operators of the form
\beq
\C{L} \supset \int d^4 \theta 2m^2 V V'. 
\eeq
In such cases we cannot use the Wess-Zumino gauge for the vector superfields; the gauge symmetry is broken. However, we can still split into ``transverse'' components in the Wess-Zumino gauge and ``longitudinal'' fields given by a chiral multiplet $S$; 
\beq
V_{massive} = V_{WZ} + \frac{1}{m\sqrt{2}}(\C{S} + \ov{\C{S}}).
\eeq
Then using notation 
\begin{align}
V_{WZ}&=\theta \sigma^\mu \ov{\theta} A_\mu + \theta^2 (\ov{\theta} \ov{\lambda}) + \ov{\theta}^2 (\theta \lambda) + \frac{1}{2} \theta^2 \ov{\theta}^2 D \nonumber \\
\C{S} &= S + \sqrt{2} \theta \psi + (\theta \theta) F_S + ...
\end{align}
we can write the mass terms as
\beq
\C{L} \supset - m^2 A_{\mu}^{\prime} A^\mu - m( \psi' \lambda + \ov{\psi}' \ov{\lambda} + \psi \lambda' + \ov{\psi} \ov{\lambda}').
\eeq
Of course, the notation above was suggestive; supersymmetric masses may be generated by axions represented by a chiral superfield $S$. For example, in a theory with $N_a$ axions $L^i = \frac{1}{\sqrt{2}} (S^i + \ov{S}^i)$ with generic K\"ahler potential $K_{ij}$, and $N_1$ $U(1)$s with coupling
\beq
\C{L} \supset \int d^4 \theta 2 M_{ij} L^i V^j,
\eeq 
the Lagrangian density becomes
\beq
\C{L} \supset \int d^4 \theta  (K^{kl} M_{ki} M_{lj}) V^i_{(m)} V^j_{(m)} 
\eeq
where now
\beq
V^i_{(m)} \equiv M^{ij} K_{jk} L^k + V^i
\eeq
is gauge invariant; under gauge transformations $V^i \rightarrow V^i + \Lambda^i + (\Lambda^i)^\dagger$, and $S^i \rightarrow S^i - \sqrt{2} K^{ij} M_{jk} \Lambda^k$.

Using the above expression, we can read off the supersymmetric  masses generated as
\beq
\int d^4 \theta  (K^{kl} M_{ki} M_{lj}) V^i_{(m)} V^j_{(m)} \supset - \frac{1}{2} (K^{kl} M_{ki} M_{lj}) A_{\mu}^i A^{j\,\mu} - M_{ij} (\psi^i \lambda^j + \ov{\psi}^i \ov{\lambda}^j ).
\eeq
However, note that the axionic fields are not in a canonical basis; if $\psi^i = U^{i}_j \psi^{\prime\,j}$ such that $K_{kl} U^k_i U^l_j = \delta_{ij}$ (and thus $U^i_k U^j_k = K^{ij}$) then the Dirac mass terms read $- M_{kj}U^k_i (\psi^{\prime\,i} \lambda^j + \ov{\psi}^{\prime\,i} \ov{\lambda}^j )$.

\section{Conclusions}

In this work we have discussed  two aspects of models with Dirac gauginos. On one side, we have described how Dirac gaugino masses are related to $U(1)$ kinetic mixing. In doing so, we have presented original formulae,  equation (\ref{new1}) and the following, for the kinetic mixing in the supersymmetric case, parallel to what was done    for the holomorphic gauge kinetic function \cite{Kaplunovsky:1994fg}. We discussed for the first time how this can used to derive the expression for Dirac gaugino masses using (\ref{new2}), for instance. In section 4, in a string D-brane example, we showed how this can make a computation simpler. 

On the other hand, we stressed that, to be soft, Dirac gaugino masses are to be associated with new interactions in the scalar sector (\ref{newinter}). These might have important phenomenological consequences, and they have not been discussed in detail in the past. We illustrated this for the simple case of moduli fields and dark matter decays.

\section{Acknowledgements}

M.~D.~G. would like to thank J\"org J\"ackel, Javier Redondo and Andreas Ringwald for many interesting discussions.

\end{document}